\documentclass[final,5p,times,twocolumn]{elsarticle}

\usepackage{latexsym}
\usepackage{amssymb, amsmath, bm, physics}
\usepackage{subcaption}
\usepackage{mathptmx}
\usepackage{flushend}
\usepackage[colorlinks,citecolor=blue,urlcolor=blue,linkcolor=blue]{hyperref}
\usepackage{aas_macros}

\journal{Physics of the Dark Universe}

\begin{document}

\begin{frontmatter}




\title{Friedmann equations of the fractal apparent horizon}

\author[1]{R. Jalalzadeh}
\ead{r.jalalzadeh@uok.ac.ir}
\address[1]{Department of Physics, Faculty of Science, University of Kurdistan,
Pasdaran Street, P.O. Box 66177-15175 Sanandaj, Iran}
\author[2,3]{S. Jalalzadeh\corref{cor1}}
\ead{shahram.jalalzadeh@ufpe.br}
\address[2]{Departamento de Fisica, Universidade Federal de Pernambuco, Recife, PE 50670-901, Brazil}
\address[3]{Center for Theoretical Physics, Khazar University, 41 Mehseti Street, AZ1096, Azerbaijan}
\cortext[cor1]{Corresponding author}

\author[4]{A. Sayahian Jahromi}
\ead{sayahian@gmail.com}
\address[4]{Zarghan Branch, Islamic Azad University, Zarghan, Iran}
\author[5]{H. Moradpour}
\ead{h.moradpour@riaam.ac.ir}
\address[5]{Research Institute for Astronomy and Astrophysics of Maragha (RIAAM), University of Maragheh, P.O. Box 55136-553, Maragheh, Iran}


\begin{abstract}
From a fractal perspective, the entropy bound of gravitational systems undergoes changes. Furthermore, in the cosmological setting, the conservation law of a perfect fluid is also altered in such systems, affecting spatial elements like volume, area, and radius. By applying the first law of thermodynamics and deriving the Friedmann equations, we can gain insight into the evolution of such a fractal cosmos. However, observations continue to necessitate the existence of a dark energy source. To address this, in this article, we have created a novel fractal $\Lambda$CDM cosmological model and determined the fractal cosmological observables. We show that the spatial fractal dimension is two, and the age of the Universe is 13.91 Gyr, by fitting the model's parameters to cosmological data.

\end{abstract}


\end{frontmatter}
\section{Introduction}

The laws of thermodynamics have shed light on the connection between gravity and thermodynamics, especially as it relates to black holes (BH) thermodynamics. Initially, the study of the thermodynamic aspects of gravity was limited to static horizons and BH thermodynamics. However, further attempts were made to explore dynamic cases. Based on these investigations, it is possible to think of the gravitational field equations as thermodynamic equations of state. 
Remarkably, thermodynamics principles may be used to produce the Friedman equations in a homogeneous and isotropic cosmology. The energy-momentum conservation law is a significant factor in this, as the first rule of thermodynamics is a conservation law \cite{Padmanabhan:2009vy}.

{If one takes a look at cosmological models with horizons, there is a resemblance to thermodynamic systems \cite{Padmanabhan:2009vy}, which makes it possible to associate entropy and temperature with them, just as with BHs. Because the laws of thermodynamics have proven to be remarkably universal, cosmological models that fulfil both the laws of thermodynamics and observational data are generally regarded as more acceptable than models that only satisfy observational tests.
}

{The equivalency between the Friedmann equation and the first law of thermodynamics on the apparent horizon in the context of general relativity, and also in the Gauss--Bonnet and Lovelock gravity theories, was originally established by Cai and Kim \cite{Cai:2005ra}. Later, this study was extended to include $f(R)$ gravity and scalar-tensor gravity theories by Akbar and Cai \cite{Akbar:2006er}. A mass-like function (with energy dimension) was later presented by Gong and Wang \cite{Gong:2007md}; this function corresponded to the Misner--Sharp--Hernandez mass on the apparent horizon. In this way, the Friedmann equations were used to derive the first law of thermodynamics.
}

{Conversely, the cosmological (the Friedmann and its extensions) equations may be obtained by extending the first law of thermodynamics to the Universe's horizon. Note that this procedure is still only a hypothesis and not a proven theorem at this time.
 However, if the correct modified entropy relation is used for each specific theory, this method seems to produce accurate findings in many cases of modified gravitational theories. By applying this method, we will be able to derive new, modified Friedmann equations that, up to a specific limit--that is, the point at which the generalised entropy assumes the standard form--include the standard equations. However, these modified equations introduce new terminology that is used in the broader context. 
 In this manner, the Tsallis entropy \cite{Lymperis:2018iuz}, the R\'enyi entropy \cite{Ghaffari:2019mrp}, the Barrow entropy \cite{Sheykhi:2021fwh}, the Sharma--Mittal entropy \cite{Kolesnichenko:2022rfj}, the Kaniadakis entropy \cite{Drepanou:2021jiv}, and the quantum deformed (q-deformed) entropy \cite{Jalalzadeh:2023mzw}, have been used to construct new modified cosmological models.
}

Finding a suitable boundary for the study of the thermodynamics of the cosmos is an essential step that requires careful consideration. The apparent horizon of the homogeneous and isotropic universe is the causal boundary that satisfies the laws of thermodynamics. This claim has been supported by many references, including \cite{cons, cons1, cons2, cons3, Cai2, CaiKimt}. Notably, the Friedmann equations result from the first law of thermodynamics being available on the apparent horizon by employing the Bekenstein entropy. Our view of the Universe is greatly affected by this relationship between thermodynamics and cosmology \cite{cons3, Cai2, CaiKimt}.

Moreover, the role of horizon entropy cannot be overstated, as it forms the foundation of investigations into thermodynamics. Thermodynamics relies significantly on entropy, and changes to the Bekenstein entropy can similarly affect the Friedmann equations. This idea has been explored in various theories that propose alternative versions of the Bekenstein entropy, as evidenced by references such as \cite{Padmanabhan:2009vy, Moradpour:2016rcy, Kom1, Kom2, Kom3, hom1, Nunes:2015xsa}. These theories offer new perspectives on the thermodynamics of the cosmos and contribute to the ongoing scientific discourse in this field.

Fractional calculus aims to expand classical calculus by including differentiation and integration principles and techniques that cover non-integer orders. In traditional calculus, differentiation and integration are limited to integer orders and do not account for non-integer or irrational orders. Fractional calculus has several applications in gravity and cosmology, which have been extensively researched. Its effectiveness in addressing gravitational forces and cosmological models makes it an essential tool for scientific inquiry \cite{Costa:2023zfg, Garcia-Aspeitia:2022uxz, Shchigolev:2021lbm, Jalalzadeh:2022uhl, Leon:2023iaq, Gonzalez:2023who, Socorro:2023ztq, Debnath2012, Momeni2012, Socorro:2023xmx}. These applications include the stochastic gravitational wave background in quantum gravity \cite{Calcagni:2020tvw}, gravitational-wave luminosity distance \cite{Calcagni:2019ngc}, CMB spectrum and inflation \cite{Rasouli:2022bug, Calcagni:2017via}, cosmology with fractional action \cite{El-Nabulsi:2012wpc, Jamil:2011uj}, discrete gravity \cite{El-Nabulsi:2013mma}, chaotic inflation with a non-minimal coupling field  \cite{El-Nabulsi:2013mwa}, phantom cosmology with a conformal coupling field \cite{Rami:2015kha}, cosmology with Ornstein--Uhlenbeck like fractional differential equation \cite{El-Nabulsi:2016dsj}, fractional action cosmology where the order parameter is variable \cite{El-Nabulsi:2017vmp}, and wormholes in fractional action cosmology \cite{El-Nabulsi:2017jss}. 

Additionally, new metrics and models of dark energy have been considered in emergent, logamediate, and intermediate models of the Universe \cite{Debnath2012}. Considering space-time is fractal, Refs. \cite{Shchigolev:2010vh, Gong:2007md} have established several exact solutions for cosmological models that do not match the standard one. Riesz's fractional derivative is used in Ref. \cite{Jalalzadeh:2022uhl} to determine the non-boundary and tunnelling wavefunctions for a closed de Sitter universe. The authors of Ref. \cite{Rasouli:2022bug} utilised fractional quantum cosmology to investigate the pre-inflation era. The analysis of the thermodynamics of fractional BHs is done in Ref. \cite{Jalalzadeh:2021gtq}. Furthermore, the Friedmann equations are modified using fractional calculus to explore the Universe's dynamics without dark energy (DE) and cold dark matter (CDM) \cite{Barrientos:2020kfp}. Furthermore, Refs. \cite{Giusti:2020rul, Torres:2020xkw, Barrientos:2020kfp, Socorro:2024poa} also discuss fractional calculus in the context of quantum cosmology and modified Newtonian dynamics (MOND).

Fractional quantum gravity is a theoretical framework that explores the possibility of describing quantum gravity in terms of fractional derivatives \cite{Jalalzadeh_2022c}. One of the intriguing concepts within this framework is the existence of fractional-fractal BHs \cite{Jalalzadeh:2021gtq}. These BHs are thought to have properties that are different from the classical BHs predicted by general relativity. As shown in Ref. \cite{Jalalzadeh_2022c}, these BHs possess a fractal structure on the apparent horizon that is characterized by self-similarity across different scales. Additionally, the entropy ($S$)-area($\mathcal A$) relationship of a fractal horizon of a fractional-fractal BH is described by $S\sim \mathcal A^\frac{d}{2}$ in which $d$ denotes the fractal dimension of the apparent horizon\cite{Jalalzadeh:2021gtq, Junior:2023fwb}. Recently, an attempt has been made to determine the corresponding Friedmann equations for the Friedmann--Lemaître--Robertson--Walker (FLRW) universe by using this entropy-area proposal and the first law of thermodynamics on the apparent horizon \cite{Coker:2023yxr}. The authors of the mentioned article utilised the ordinary energy-momentum tensor and the standard volume, area, and radius to handle their calculations. 

It is worth noting that, despite the mathematical similarity of fractional-fractal entropy of BHs to that of the Tsallis and Barrow entropies \cite{Tsallis:2012js, Barrow:2020tzx}, they originated from different physics \cite{Rasouli:2022bug, Jalalzadeh:2022uhl, Jalalzadeh:2021gtq, Junior:2023fwb, Jalalzadeh:2023mzw, Costa:2023zfg}. Moreover, the continuity equation of a perfect fluid in fractional-fractal cosmology is different from the conservation equation for ordinary perfect fluid \cite{Junior:2023fwb}. Additionally, the proper volume, horizon area, and radius in fractional-fractal cosmology stored the effects of fractionality and rescaled compared to their ordinary counterparts \cite{Rasouli:2022bug, Jalalzadeh:2022uhl, Jalalzadeh:2021gtq, Junior:2023fwb, Jalalzadeh:2023mzw, Costa:2023zfg}.

Henceforth, the foremost objective of this article is to ascertain the modified Friedmann equations by employing fractal entropy and the fractal extension of the conservation law of matter fields, along with the fractal considerations on the apparent horizon's volume, area, and radius.

The following is an outline of this article. In Section \ref{sec2}, we apply the first law of thermodynamics of FLRW cosmology, Cai--Kim temperature, and fractal continuity equation to the fractal volume and surface of the apparent horizon to obtain a fractal extension of the Friedmann equations. In Section \ref{sec3}, we use the obtained fractal Friedmann equations to analyse the corresponding cosmology. We obtain the fractal deceleration parameter, the age of the Universe, and the transition redshift. By defining fractal density parameters, we compare our equations with cosmological data in section \ref{fitting}. We show that the Universe's age can considerably surpass the customary standard flat $\Lambda$CDM model.  In the final section, we draw our conclusions.

\section{Fractal Friedmann equations}\label{sec2}

Numerous studies and references, such as \cite{Cai2, cons3, DiCriscienzo:2007pcr}, have been dedicated to exploring the intricacies of the thermodynamics of FLRW spaces. To comprehensively understand the FLRW apparent horizon's thermodynamical properties, one can refer to a comprehensive review presented in Ref. \cite{DiCriscienzo:2007pcr}.

The line element of the FLRW universe is given by
\begin{eqnarray}\label{1}
ds^{2}=-dt^{2}+a^{2}\left( t\right) \left( \frac{dr^{2}}{1-k
r^{2}} +r^{2}d\Omega^{2}\right),
\end{eqnarray}
 where the curvature constant $k=-1,0,1$ addresses
the open, flat, and closed universes, respectively. The coordinate $r$ that moves with the Universe's expansion is not equivalent to the areal radius. Instead, the areal radius (proper radius) depends on the time coordinate, $t$, and is given by $\tilde r=a(t)r$. The apparent
horizon (as the proper casual boundary) is located at \cite{cons,
cons1, cons2, cons3, Cai2, CaiKimt}
\begin{eqnarray}\label{ah}
\tilde{r}_A=a(t)r_A=\frac{1}{\sqrt{H^2+\frac{k}{a(t)^2}}}.
\end{eqnarray}
Indeed, $r_A$ is a marginally trapped surface, and it is calculated by solving
\begin{eqnarray}\label{ah2}
\partial_{\alpha}\tilde r\partial^{\alpha}\tilde r=0.
\end{eqnarray}

In fractal cosmology, the effective area of the apparent horizon has a random fractal structure \cite{Jalalzadeh:2022uhl, Junior:2023fwb}. As a result, the effective areal radius, $R$, the corresponding surface area $\mathcal A$, and volume, $\mathcal V$ are given by \cite{Junior:2023fwb}
\begin{equation}
    \label{P6}
    \begin{split}
    & R=(4\pi)^\frac{2-\alpha}{4\alpha}\left(\frac{\tilde{r}_A}{L}\right)^\frac{2+\alpha}{2\alpha}L,\\
  &\mathcal A=4\pi R^2,\\
   &\mathcal V=\frac{4\pi}{3}R^3,
\end{split}
\end{equation}
where $\alpha$ is Lévy’s fractional parameter \cite{Laskin:1999tf}, and $L$ is a fractional length scale that must be introduced to maintain dimensional consistency. Notably, this scale corresponds to the grid size of the fractal structure \cite{Junior:2023fwb}. Also, if we assume the matter content of the model universe is a perfect fluid with energy density $\rho$ and pressure $p$. As a result, the Misner--Sharp--Hernandez mass of a sphere of radius $R$ is
\begin{equation}
    \label{Mass}
    M_\text{MSH}=\rho\mathcal V=M_0\left(\frac{\tilde{r}_A}{L}\right)^\frac{6+3\alpha}{2\alpha},
\end{equation}
where $M_0$ is the grid mass, given by
\begin{equation}
    \label{Mass0}
    M_0=\frac{(4\pi)^{\frac{6+\alpha}{4\alpha}}}{3}\rho L^3.
\end{equation}
Note that for $\alpha=2$, these definitions reduce to their ordinary counterparts in the standard model of cosmology.

It is vital to have a better comprehension of the subject at hand, and this can only be done by realising the relevance of Eq. (\ref{Mass}). 
 Let us take a closer look at its meaning. Suppose that within a volume of size $l$, we can locate $N_0$ particles with a mass of $m_0$. We find $N_1 = k_NN_0$ particles if we increase the volume to a bigger size of $R_1 = k_rl$.
 In a structure that exhibits self-similarity, the values of $k_r$ and $k_N$ remain consistent even when the scale is altered. Therefore, in a structure with a size of $R_n=k_r^nl$, we would observe $N_n = k_N^nN_0$ particles. Consequently, we can express the relationship between the number $N$ (or mass $M = m_0N$) and the length $l$ in the form $N(R)=CR^D$ or $M(R)=m_0N(R)=Cm_0R^D$, where $D$ denotes the fractal dimension, depends on the rescaling factors $k_r$ and $k_N$, and $C=N_0/r_0^D$ gives the prefactor $C$.

The argument presented above, along with Eq. (\ref{Mass}), demonstrates that the fractal dimension of the fractal cosmology is linked to L\'evy's fractional parameter. Specifically, the relationship is given by the equation
\begin{equation}
    \label{dimension}
    D=\frac{3}{2}\left(1+\frac{2}{\alpha}\right).
\end{equation}
Note that when $\alpha=2$, $D$ equals the ordinary dimension of three.
It is reasonable to suppose that the grid size in the fractional BH \cite{Jalalzadeh:2021gtq} is equal to the Planck length $L=L_\text{Pl}$.
 The Planck length became less important in later phases of cosmology, and the grid size was determined by the size of cosmic dust distributed in a fractal pattern. Therefore, if we assume that we can apply Eqs. (\ref{P6}) in fractal cosmology, we can relax the value of $D$ from $\alpha$ and consider it as an independent parameter. When the dimension is less than 3, we refer to it as a fractal structure. A linear distribution of matter occurs when $D=1$, surface distribution when $D=2$, and space-filling distribution when $D=3$ \cite{Mureika:2006tz}.

 On the apparent horizon of the FLRW cosmological model, the first law of thermodynamics is expressed as \cite{Cai2, CaiKimt}
\begin{eqnarray}\label{4}
dE=\mathcal A\Psi+Wd\mathcal{V},
\end{eqnarray}
 where $W=-\frac{1}{2}h_{ab}T^{ab}=\frac{\rho-p}{2}$ is the work density and
\begin{equation}\label{5}
\mathcal A\Psi=\mathcal A\underbrace{(T_a^b \partial_b \tilde{r}_A+W\partial_a \tilde{r}_A)}_{\psi_a}dx^a.
\end{equation}
Here, $\psi_a$ denotes the projection of $T_{\mu\nu}$ on
the direction normal to the boundary. Considering that the Cai--Kim approach \cite{CaiKimt, Debnath:2020inx} relies on the fundamental assumption, and since $dE=\rho d\mathcal{V}+\mathcal{V}d\rho$, one can use Eq.~(\ref{5}) or directly calculate $dE-Wd\mathcal{V}$ to obtain
\begin{eqnarray}\label{6}
\mathcal A\Psi=dE-Wd\mathcal{V}=\mathcal{V}d\rho.
\end{eqnarray}

Moreover, the common result (based on using ordinary non-fractional spatial elements)
\cite{Moradpour:2016rcy} is achieved if $\alpha=2$. The entropy of a
fractal horizon is reported as \cite{Jalalzadeh:2021gtq}
\begin{eqnarray}\label{7}
S_A= \frac{\mathcal A}{4G},
\end{eqnarray}
and obviously, $\alpha=2$ recovers the Bekenstein entropy, a
desired result. For the Cai--Kim temperature of
the horizon (which is, in fact, the corresponding Hawking temperature of the apparent horizon), calculations lead to
\cite{CaiKimt}
\begin{eqnarray}\label{3}
T=\frac{1}{2\pi\tilde{r}_A}.
\end{eqnarray}

 As it has previously been noted, the area and volume of a fractional-fractal system differ from those of a non-fractal one because the radius is also modified under the shadow of fractionality as given by Eq. (\ref{P6}) $\tilde r_A\rightarrow R$ \cite{Junior:2023fwb}. The latter motivates us to modify the Cai--Kim temperature as 
\begin{eqnarray}\label{30}
T^\text{(fractal)}=\frac{1}{2\pi R}.
\end{eqnarray}
It is beneficial to note here that the radius temperature dependence of Hawking radiation of BHs experiences the same modification in the presence of fractionality \cite{Junior:2023fwb}. Finally, since $T^\text{(fractal)}dS_A=-\mathcal A\Psi$ \cite{CaiKimt}, one reaches
\begin{eqnarray}\label{8}
T^\text{(fractal)}dS_A=-\mathcal{V} d\rho,
\end{eqnarray}
{leading (after integrating) to the following equation
 \begin{equation}
     \label{9aa}
     \frac{(4\pi)^{1-\frac{D}{3}}}{L^2}\left(\frac{L}{\tilde r_A}\right)^\frac{2D}{3}=\frac{8\pi G}{3}\rho
,
\end{equation}
where $D$ is defined by (\ref{dimension}).
The modified first Friedmann equation can be obtained by rearranging the previous equation and substituting the definition of the effective areal radius provided in (\ref{P6})
}
\begin{eqnarray}\label{9}
H^2+\frac{k}{a(t)^2}=\left(\frac{2GL^2\rho}{3} \right)^{\frac{D}{3}-1}\frac{8\pi G}{3}\rho.
\end{eqnarray}

 Combining the time derivative of this equation with the corresponding fractional-fractal continuity equation, i.e., \cite{Jalalzadeh:2021gtq}
\begin{eqnarray}\label{10}
\dot{\rho}+DH(\rho+p)=0,
\end{eqnarray}
one is finally able to get the second Friedmann equation as
\begin{eqnarray}\label{11}
\frac{\ddot a}{a}=-\left(\frac{2GL^2\rho}{3} \right)^{\frac{D}{3}-1}\frac{4\pi G}{3}(\rho+p).
\end{eqnarray}

 It is easy to check that for $D=3$, the prefactor $\left(\frac{2GL^2\rho}{3} \right)^{\frac{D}{3}-1}$ in fractal Friedmann equations (\ref{9}) and (\ref{11}) reduce to unity, and we 
recover the standard Friedmann equations. Note that the difference between fractal Friedmann equations obtained here with
the results of Ref.~\cite{Coker:2023yxr} is due to $i$)
fractal volume and surface area are considered here, $ii$) the re-scaled Cai--Kim temperature is taken into account (of course, it only affects the coefficient $D$), and $iii$) the fractional-fractal continuity equation~(\ref{5}) is used instead of the
ordinary continuity equation employed in Ref.~\cite{Coker:2023yxr}.
{It should also be noted that, as demonstrated by Eqs. (\ref{9aa}) and (\ref{10}), the fractal structure of the apparent horizon has impacted not only the geometric aspect (the left-hand side) of the Friedman equation but also the distribution of matter. Thus, both components are subject to modification.}

\section{Fractal FLRW Cosmology
}\label{sec3}

The cosmic fluid is composed of three non-interacting components: the radiation (which includes relativistic particles) with energy density $\rho_\text{(rad)}$, the cosmic dust (constituted of baryonic and cold dark matter) with an energy density $\rho_\text{(dust)}$, and the vacuum energy (or the cosmological constant) in which $\rho_{(\Lambda)}=\frac{\Lambda}{8\pi G}$,  where $\Lambda$ is the cosmological constant. Therefore
\begin{equation}
    \label{3-1}
\rho=\sum_i\rho_\text{(i)}=\rho_\text{(rad)}+\rho_\text{(dust)}+\rho_{(\Lambda)}.
\end{equation}
These components add to the Universe's overall energy density; radiation and cosmic dust were the main contributors in the early stages, but as the Universe expands, vacuum energy becomes more dominant.

Regarding the fractal Friedmann equation (\ref{9}), the density parameters' fractional-fractal extension can be defined similarly to the standard model of cosmology
\begin{equation}
    \label{3-2}
    \tilde\Omega^\text{(i)}=\left(\frac{L^2H^2}{4\pi} \right)^{1-\frac{D}{3}}\frac{8\pi G\rho_\text{(i)}}{3H^2}=\left(\frac{L^2H^2}{4\pi} \right)^{1-\frac{D}{3}}\Omega^\text{(i)},
\end{equation}
where $\Omega^\text{(i)}=\frac{8\pi G\rho_\text{(i)}}{3H^2}$ denote the usual density parameters of the standard model of cosmology. The fractional-fractal density parameters closely resemble those in the standard model of cosmology. However, they are modified by a prefactor $\left(\frac{L^2H^2}{4\pi} \right)^{1-\frac{D}{3}}$. This prefactor takes into account the effects of grid size $L$ and the fractal dimension $D$. Utilising these fractal extensions of density parameters, the fractal Friedmann (\ref{9}) takes the following
 form
\begin{equation}
    \label{3-3}
    \sum_i\tilde\Omega^\text{(i)}=\left(1-\Omega^{(k)}\right)^\frac{D}{3},~~~~\text{i}=\text{rad, dust}, \Lambda,
\end{equation}
where $\Omega^{(k)}=-\frac{k}{a^2H^2}$, is the standard curvature density parameter.

Also, the second fractal Friedmann equation (\ref{11}) gives us the deceleration parameter
\begin{equation}
    \label{3-4}
    q=-\frac{\ddot a}{aH^2}=\frac{1}{2}\left(1-\Omega^{(k)}\right)^\frac{D}{3}\sum_i(1+3\omega_i)\tilde\Omega^\text{(i)}.
\end{equation}
Note that $\omega_i$ is the equation of state parameter of the i-component of the cosmic fluid given by $\omega_i=\frac{p_\text{(i)}}{\rho_\text{(i)}}$, where $p_\text{(i)}$ is the pressure of the i-component of the cosmic fluid.

In addition, employing the continuity equation (\ref{10}), the fractional-fractal Eqs. (\ref{3-3}) and (\ref{3-4})
can be rewritten as
\begin{equation}
    \label{3-5}
    \left(\frac{H}{H_0}\right)^2=\frac{\Omega^{(k)}_0}{a^2}+\left(\sum_i\tilde\Omega^\text{(i)}_0a^{-D(1+\omega_i)} \right)^\frac{3}{D}:=E^2,
\end{equation}
and
\begin{equation}
    \label{3-6}
    q=\frac{1}{2E^\frac{2D}{3}}\left(1-\frac{\Omega_0^{(k)}}{E^2a^2} \right)^{1-\frac{D}{3}}\sum_i(1+3\omega_i)\frac{\tilde\Omega_0^{(i)}}{a^{D(1+\omega_i)}},
\end{equation}
where subscript zero denotes the value of the corresponding quantity at the present epoch. {Utilising Eq. (\ref{3-5}), one can easily find the age of the Universe as
\begin{equation}
    \label{age}
    t_0=
    \frac{1}{H_0}\displaystyle\int_0^1\frac{da}{a\left[\Omega_0^\text{(k)}a^{-2}+\left(\displaystyle\sum_i\tilde\Omega_0^\text{(i)}a^{-D(1+\omega_i)} \right)^\frac{3}{D} \right]^\frac{1}{2}}.
\end{equation}
}

Note that based on the age-dating of field halo stars and GCs, it is strongly suggested that the oldest stars in the Milky Way have ages that are comparable to the age of the Universe in the standard $\Lambda$CDM model of cosmology \cite{Verde:2013wza}. For instance, the age of the Population II halo in the Milky Way, characterized by a deficiency in metals and a high velocity, is estimated to be $14.46\pm 0.31$ Gyr, according to Bond et al. \cite{Bond}. Also, utilizing new stellar models, VandenBerg et al. \cite{VandenBerg} estimate the age to be $14.27 \pm 0.80$ Gyr. These estimations of the star's age slightly surpass ($\sim2\sigma$) the age of the Universe, based on CMB Planck18 data.

{On the other hand, fractal cosmology has the potential to surpass the estimated age of the Universe in the standard model of cosmology to overcome the above difficulty in the explanation of age dating of field halo stars and GCs. For example, let us consider a flat ($k=0$ or $\Omega^{(k)}=0$) fractal cosmological model, and utilise the above age equation to compare the dimensionless age of the Universe, $t_0H_0$, for three values of the fractal dimension $D=1,2$ and 3. Fig. \ref{sAge} shows the age of a fractal universe in units of Hubble time versus $\tilde\Omega_0^\text{(dust)}$.
\begin{figure}[h!]
  \centering
  \includegraphics[width=9cm]{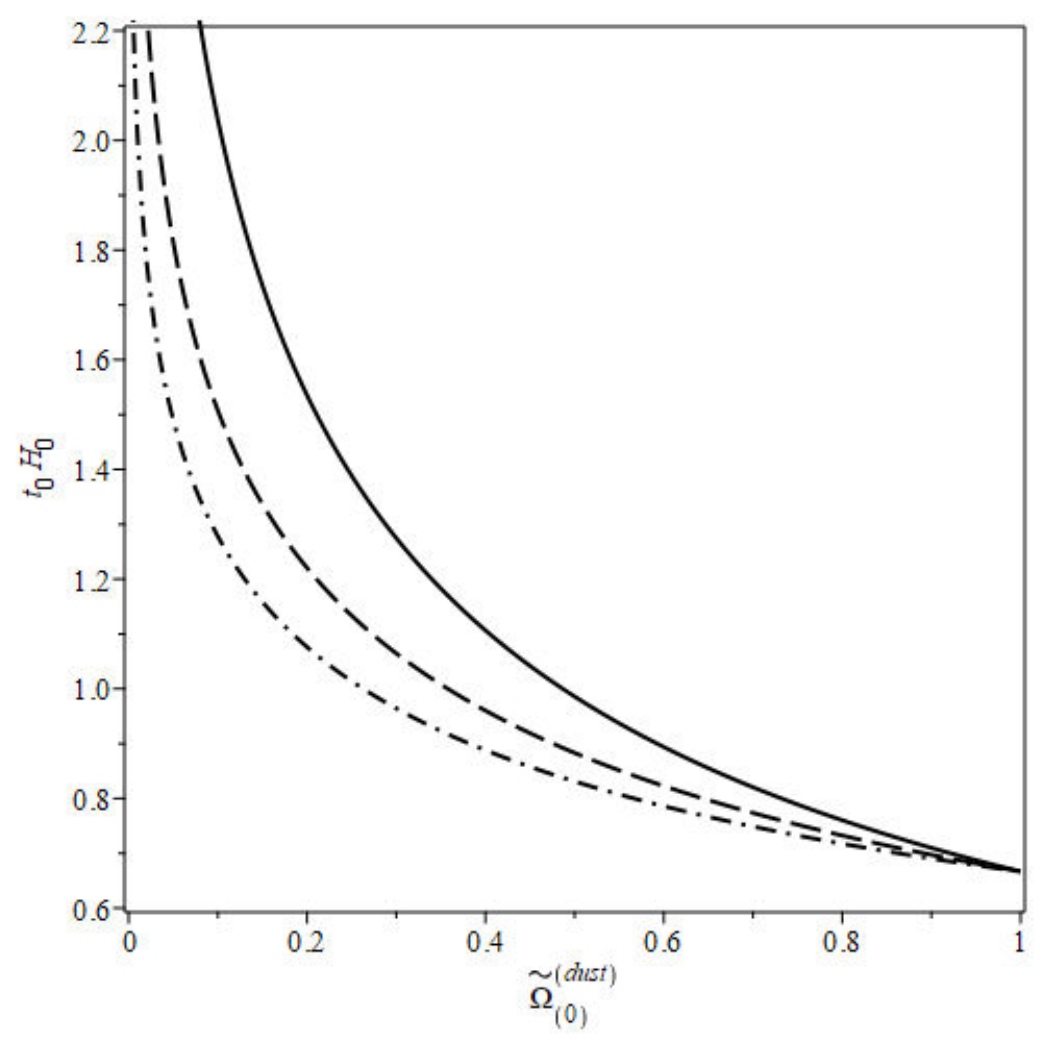}
  \caption{\small The age of the Universe as a function of $\tilde\Omega_0^\text{(dust)}$. The line, dashed line, and dash-doted line show the age for $D=1,2$ and 3, respectively. }\label{sAge}
\end{figure}
As we can see in Fig. \ref{sAge}, the lowest and greatest values of the age are provided for $D=3$ and $D=1$, respectively, for a given value of $\tilde\Omega_0^\text{(dust)}$. This demonstrates that the standard model of cosmology provides a smaller estimate of the age of the Universe for a given value of the dust density parameter at the present epoch.}

\section{Data-fitting}\label{fitting}

To have a better grasp of the influence of the fractal dimension on cosmic dynamics, it is helpful to examine the simple flat ($k=0$ or $\Omega^{(k)}=0$) universe model with cosmic dust and the cosmological constant.

{Our objective, then, is to use the Markov Chain Monte Carlo (MCMC) approach to restrict the parameters of the fractal FLRW model at the background level using a recent dataset of SNeIa (Pantheon+) and Observational Hubble Data (OHD).
 The fitting method employed in this approach is the maximum likelihood estimation. Following this method, the total likelihood function, denoted as $\mathcal{L}_\text{tot} \backsimeq e^{-\frac{\chi^2_\text{tot}}{2}}$, is maximized by minimizing $\chi^2_\text{tot}$. For determining $\chi^2_\text{tot}$, the following observational dataset is utilised: the Hubble evolution data ($H(z)$, 37 points) \cite{Yu:2017iju}, (where $z=1/a-1$ is the redshift) and type Ia supernova data (SNIa, 1701 points from Pantheon$+$ sample) \cite{Scolnic:2021amr}. Consequently, the $\chi^2_\text{tot}$ for the initial analysis can be written as
\begin{equation}\label{1RRR}
    \chi^2_{\text{tot}}(p)=\chi^2_{\text{OHD}}(p)+\chi^2_{\text{SN}}(p),
\end{equation}
where the statistical vector $p$ comprises free parameters of the model $(\tilde\Omega^\text{(dust)}_0, H_0, D)$.}

Here, we used the Hubble parameter data from Table 1 of Ref. \cite{Yu:2017iju} (obtained by the BAO signal in the distribution of galaxies, the BAO signal in the distribution of the Ly$\alpha$ forest alone, and the cross-correlation with QSOs), {a standardised compilation of 37 data points collected from the recently accurate estimates $H(z)$ within the redshift range of $0.07 \leq z \leq 2.36$ to obtain the Hubble parameter data. The chi-square, $(\chi^2)$, for the OHD is calculated as follows
\begin{equation}\label{2RRR}
 \chi^2_{\text{OHD}}(p)=\sum_{i=1} ^{37}\frac{\left[H_{\text{th}}(z_i,p)-H_{\text{ob}}(z_i)\right]^2}{\sigma^2_i},   
\end{equation}
the term $\sigma_i$ represents the Gaussian error associated with the measured value of $H_i$.}

{In Table \ref{table3}, we presented the outcomes of our statistical analysis for a flat $\Lambda${CDM} and the fractal $\Lambda${CDM} through the utilization of OHD data. As it is evident, both models are consistent with the observational data.}

\begin{table}[h!]
    \centering
    \begin{tabular}{|c|c|c|}
    \hline
 Parameter &  $\Lambda${CDM}  &    Fractal $\Lambda${CDM}\\
\hline
{\boldmath$\tilde\Omega^\text{(dust)}_0 $}  & $0.258^{+0.056}_{-0.046} $   &  $0.37^{+0.23}_{-0.16}   $ \\
{\boldmath$H_0            $} & $70.1^{+4.2}_{-4.3}     $ & $67.4^{+5.6}_{-6.7}       $\\
{\boldmath$D             $} & $3$    & $2.2^{+1.1}_{-1.4}      $\\
{\boldmath$\chi^2_\text{min}   $} & $18.08$& $18.08 $\\
\hline
\end{tabular}
    \caption{The best fitting values of free parameters of the model, and the minimal value of $\chi^2_{OHD}$ for the fractal model as well as for the flat $\Lambda$CDM using the $H(z)$ dataset.}
    \label{table3}
\end{table}

Type SNIa observations are essential for gathering essential data that contributes to a complete picture of the expansion of the Universe. The main evidence in favour of the notion of the Universe's accelerated expansion comes from these findings.
 To extract the most accurate and reliable results from the SNIa data, it is crucial to carefully compare the observed distance modulus of SNIa detection with the corresponding theoretical value.
 In light of this, we have employed {the Pantheon+ sample, a recent dataset of SNeIa, consisting of $1701$ data points of distance modulus $\mu_\text{obs}$ at various redshifts $z_i$ within the range $0.001 < z_i < 2.26$. The corresponding chi-squared statistic, $\chi^{2}$, is defined based on the comparison between the observed and theoretical distance modulus as follows
\begin{equation}\label{3RRR}
    \chi^2_{\text{sn}}(p)=\sum_{i=1}^{1701}\frac{[\mu_{\text{th}}(z_i,p)-\mu_{\text{ob}}(z_i)]^2}{\sigma^2_i},
\end{equation}
and
\begin{multline}
    \mu_{\text{th}}(z_i,p)=\\5\log_{10}{\Big[(1+z)\int^z _0\frac{dx}{E(x)}\Big]}+42.384-5\log_{10}h-19.37.
\end{multline}
Table \ref{table4} presents the parameter values obtained through constraining the parameters of the $\Lambda$CDM and Fractal$\Lambda$CDM by pantheon$+$ data.}
\begin{table}[h!]
    \centering
    \begin{tabular}{|c|c|c|}
    \hline
 Parameter &  $\Lambda ${CDM} &    Fractal $\Lambda ${CDM}\\
\hline
{\boldmath$\tilde\Omega^\text{(dust)}_0 $}  & $0.352^{+0.054}_{-0.045}   $   &  $0.393^{+0.073}_{-0.070}  $ \\
{\boldmath$H_0            $} & $69.14^{+0.61}_{-0.69}      $ & $68.98^{+0.70}_{-0.70}      $\\
{\boldmath$D             $} & $3$    & $1.99^{+0.99}_{-0.98}        $\\
{\boldmath$\chi^2_\text{min}   $} & $758.58$& $758.06$\\
\hline
\end{tabular}
    \caption{The best fitting values of free parameters of the model, and the minimal value of $\chi^2_{SN}$ for the fractal $\Lambda$CDM and flat $\Lambda$CDM using the SNe dataset. The units of $H_0$ is $\text{Km}\cdot \text{s}^{-1}\cdot\text{Mpc}^{-1}$.}
    \label{table4}
\end{table}

{Finally, we perform an overall statistical analysis using joint analyses of SNe$+$H(z) by
the MCMC algorithm. Figure \ref{fig1RRR} displays the best-fit values for the free parameters of the fractal $\Lambda$CDM model, i.e., $(\tilde\Omega^{\text{dust}}_0, H_0, D)$ with $68\%$ and $95\%$ confidence level contours, and Fig. \ref{fig2RRR} includes the comparison of the fractal $\Lambda$CDM model with the flat $\Lambda$CDM standard model of cosmology.
\begin{figure}[ht]
  \centering
  \includegraphics[width=8cm]{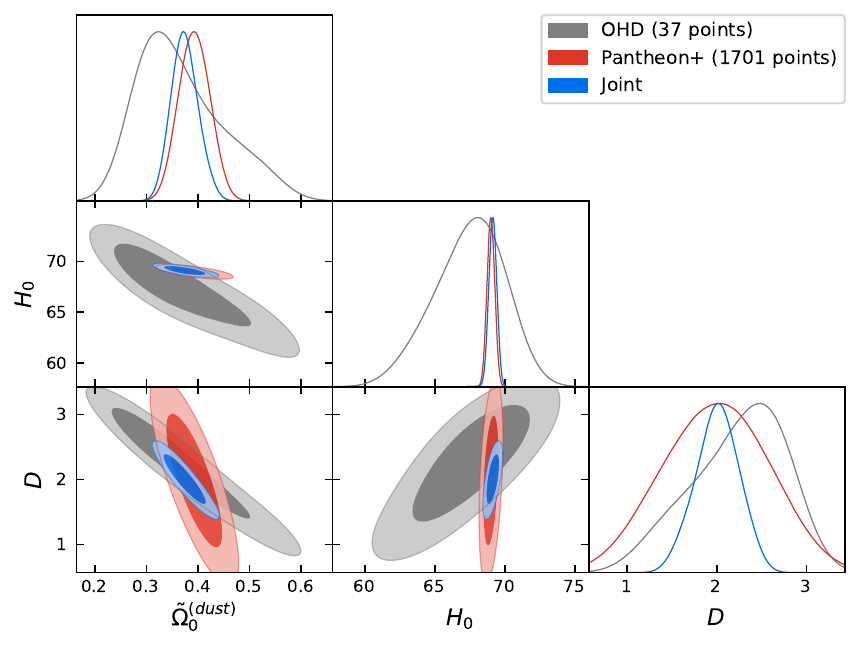}
  \caption{\small The $\%68$ and $\%95$ confidence level contours for the fractal $\Lambda$CDM model using the OHD, pantheon$+$, and joint datasets.}\label{fig1RRR}
\end{figure}
\begin{figure}[ht]
  \centering
  \includegraphics[width=7cm]{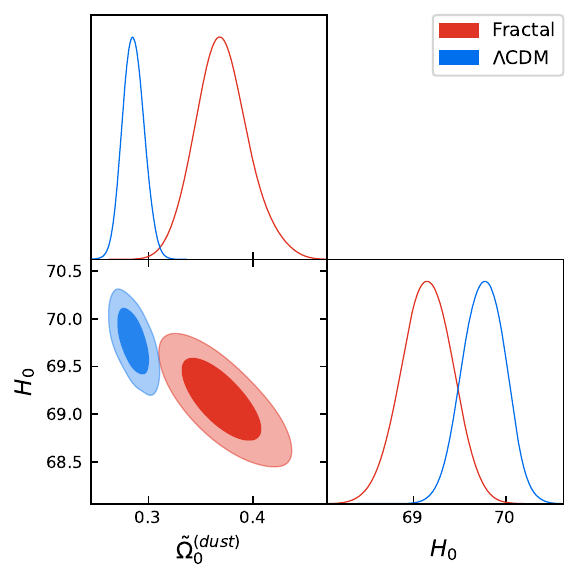}
  \caption{\small The $\%68$ and $\%95$ confidence level contours for the fractal $\Lambda$CDM and $\Lambda$CDM models using the SNe$+$H(z) joint dataset}\label{fig2RRR}
\end{figure}}

{The best-fit range (\%99 confidence level) from observational data analysis for the free parameters of the fractal model and the $\Lambda$CDM model is available in Table \ref{table1}.
\begin{table}[h!]
    \centering
    \begin{tabular}{|c|c|c|}
    \hline
 Parameter &  $\Lambda${CDM}  &    Fractal $\Lambda${CDM}\\
\hline
{$\tilde\Omega^\text{(dust)}_0 $}  & $0.285^{+0.026}_{-0.025}  $   &  $0.375^{+0.067}_{-0.067}  $ \\
$H_0            $ & $69.76^{+0.56}_{-0.56}     $ & $69.11^{+0.71}_{-0.71}     $\\
$D             $ & $3$    & $2.00^{+0.67}_{-0.61}      $\\
$\chi^2_\text{min}$    & $796.53 $& $781.23$\\
\hline
\end{tabular}
    \caption{The best fitting values of free parameters of the model, and the minimal value of $\chi^2_{tot}=\chi^2_{OHD}+\chi^2_{SN}$ for the fractal model as well as for the flat $\Lambda$CDM using the SNe+$H(z)$ joint dataset. The units of $H_0$ is $\text{Km}\cdot \text{s}^{-1}\cdot\text{Mpc}^{-1}$.}
    \label{table1}
\end{table}}

\begin{figure}[h!]
  \centering
  \includegraphics[width=9cm]{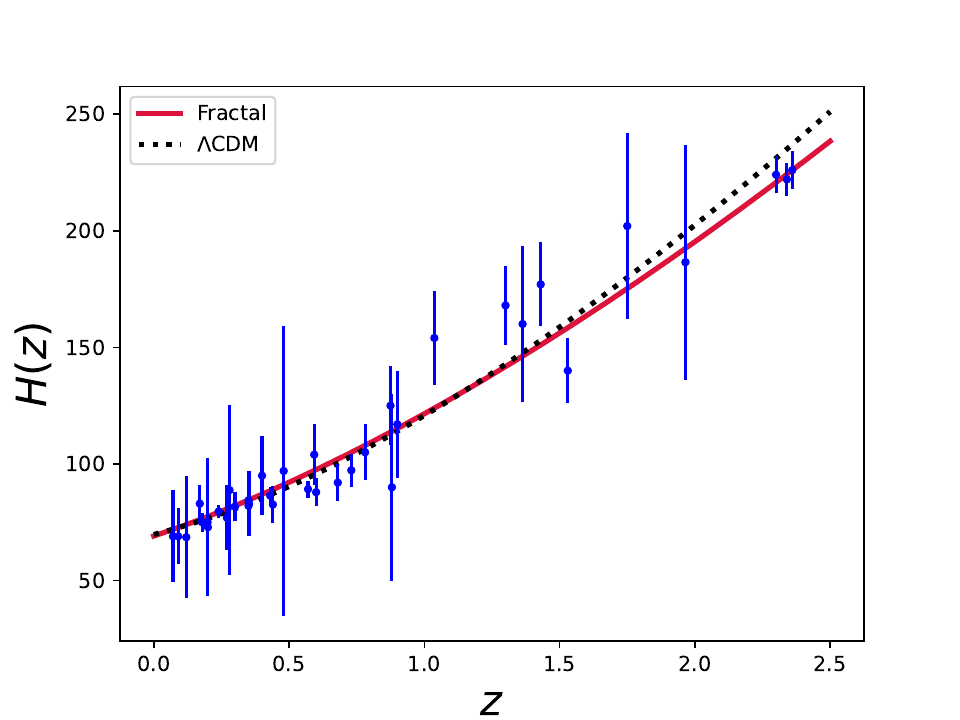}
  \caption{\small Prediction of the $H(z)$ (in units $\text{Km}\cdot \text{s}^{-1}\cdot\text{Mpc}^{-1}$) versus redshift z with error bars.  The red-solid line denotes the fractal$\Lambda$CDM model. The dark blue points denote the prediction of the $\Lambda$CDM..}\label{fig3RRR}
\end{figure}

\begin{figure}[h!]
  \centering
  \includegraphics[width=9cm]{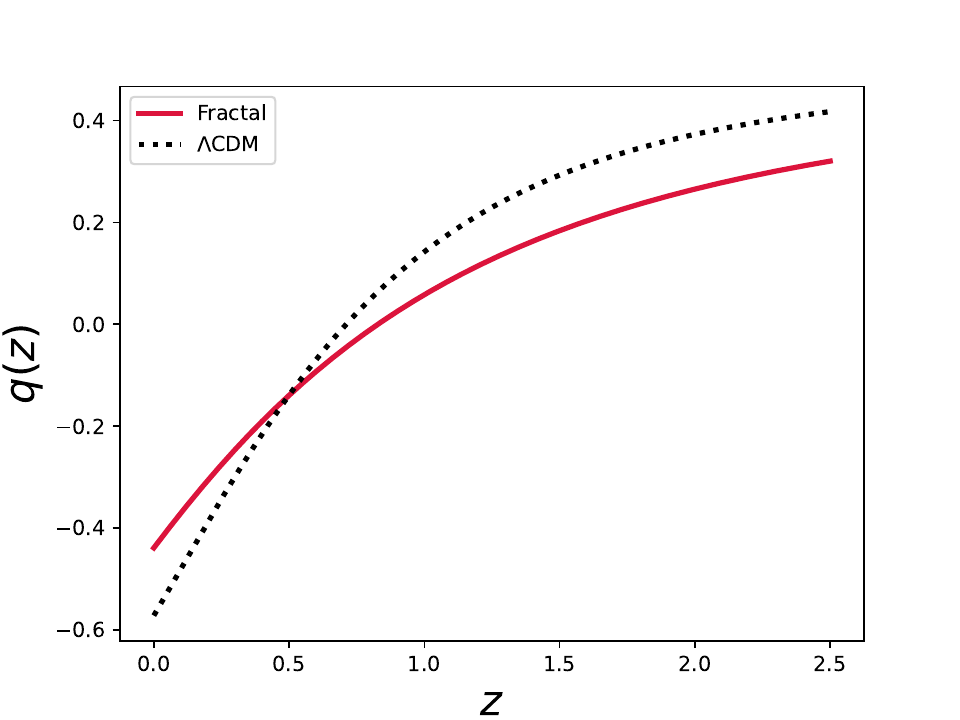}
  \caption{\small Prediction of the deceleration parameter $q(z)$ (red-solid line) using the best-fit fractal model is compared to the prediction of the $\Lambda$CDM (black dotted line).}\label{fig4RRR}
\end{figure}

\begin{figure}[h!]
  \centering
  \includegraphics[width=9cm]{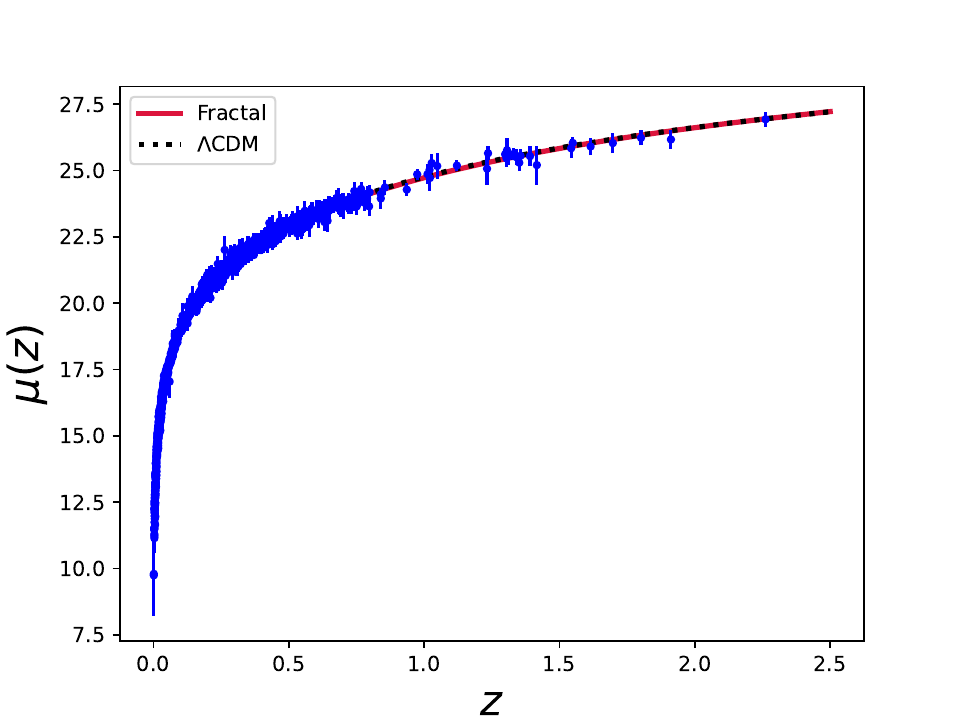}
  \caption{\small Prediction of the $\mu(z)$ (red-solid line) using the best-fit fractional $\Lambda$CDM model is compared to the observational data (dark blue points -with vertical lines indicating the uncertainty-) and the prediction of the $\Lambda$CDM (black dotted line). }\label{fig5RRR}
\end{figure}

{Using the constraint values of the model parameters from the Pantheon+ and Hubble joint dataset, we next compared our model parameterization with the $\Lambda$CDM model by studying the evolution of the Hubble parameter $H(z)$, the deceleration parameter $q(z)$, and the distance modulus $\mu(z)$.
 Our analysis is illustrated in Figs \ref{fig3RRR}, \ref{fig4RRR}, and \ref{fig5RRR}.} Fig. \ref{fig3RRR} showcases the relationship between the Hubble parameter $H$ and the redshift $z=\frac{1}{a}-1$. The Pantheon Survey is visually depicted in Fig. \ref{fig5RRR}, which showcases the standard Hubble diagram of SNIa, graphically presenting the relationship between the observed distance modulus and the corresponding redshift. Furthermore, Fig. \ref{fig5RRR} also highlights the comparison between our proposed model and the widely accepted $\Lambda$CDM model, revealing intriguing and significant similarities which further strengthen the empirical evidence and support the validity of our findings. In addition, Fig. \ref{fig4RRR} shows the behaviour of the deceleration parameter concerning redshift for the best-fit values in Table \ref{table1}.
{The parameterization of our model closely matches the observed data. The cosmological parameters of the fractal model and the flat $\Lambda$CDM at the present time ($z=0$), the Universe's age, and the corresponding transition redshift ($z_t$) are summarised in Table \ref{table2}.} 
It should be emphasised that both the fractal $\Lambda$CDM and the conventional flat $\Lambda$CDM models agree with the latest cosmological findings, concerning their respective estimates of the transition redshift $z_t$ and $q_0$.

\begin{table}[h!]
    \centering
    \begin{tabular}{|c|c|c|}
    \hline
 Parameter &  $\Lambda${CDM}  &    Fractal $\Lambda$CDM \\
\hline
{\boldmath$q_0   $} & $-0.572176^{+0.03}_{-0.03}  $  & $-0.440145^{+0.09}_{-0.09}$\\
{\boldmath$t_0           $} & $  13.7 ^{+0.46}_{-0.43}   $ & $ 13.9105 ^{+0.151}_{-0.152}   $\\
{\boldmath$\tilde\Omega^{(\Lambda)}_0             $} & $0.714814^{+0.02}_{-0.02}$    & $0.626793 ^{+0.06}_{-0.06}   $\\
{\boldmath$z_t   $} & $ 0.71^{+0.07}_{-0.07} $ & $0.82^{+0.151}_{-0.155}$ \\
\hline
\end{tabular}
    \caption{present values obtained for the cosmological parameters, age of the Universe, and the transition redshift using best-fit values of the free parameters for the fractal $\Lambda$CDM and flat $\Lambda$CDM models.}
    \label{table2}
\end{table}


{It is noteworthy to observe that the model examined in this paper posits that the fractal dimension $D$ is equivalent to 2. This outcome aligns with the findings of numerous redshift surveys, which have consistently reported a fractal dimension of approximately $D=2$ for the clustering of galaxies on a large scale \cite{SylosLabini:1997jg,Baryshev:2004mh,Yadav:2005vv,2011ARep324V,Capozziello:2009qu}.} These results strengthen the validity of our findings. According to Table (\ref{table2}) we obtain the age of the fractal Universe $13.910\pm 0.15$ Gyr. This demonstrates that the temporal extent of a fractal universe has the potential to considerably surpass the customary standard flat $\Lambda$CDM model of cosmology, wherein the age of the Universe is estimated to be $13.7\pm 0.4$Gyr in Table \ref{table2}, and $13.800\pm 0.024$ Gyr according to the Planck18 data \cite{Planck:2018vyg}.  The concept of fractal cosmology seems to possess the inherent capacity to offer a more comprehensive and improved elucidation for the aforementioned observations. Our model could be considered a viable alternative to $\Lambda$CDM through these findings. 

\section{Conclusions}

According to fractional quantum gravity, the event horizon of a BH possesses the intriguing characteristic of being a surface that exhibits a random fractal pattern, meaning it displays intricate and self-similar patterns at various scales. Moreover, it is interesting to note that this property of a fractal horizon extends beyond BHs and also applies to the event horizon of de Sitter space-time \cite{Jalalzadeh:2022uhl}. This implies that the event horizon of de Sitter space-time shares the same fascinating property of a random fractal surface. These insightful findings and ideas build upon the notion of a fractal horizon and successfully incorporate it into the broader cosmic framework. In light of this, the Friedmann equations are derived from the extension of the fractal entropy-area relationship of a fractional BH into the context of the apparent horizon of an isotropic and homogeneous universe. 

This was accomplished by employing the FLRW universe's fractal apparent horizon in conjunction with the first law of thermodynamics. Also, considering the differences between fractal and non-fractional cosmological models summarized as
\begin{itemize}
 \item The effective proper distance is re-scaled, affecting the corresponding volume and the horizon's area.
    \item Regarding the effective proper radius and proper volume, the energy-momentum conservation law of a perfect fluid is modified in fractal cosmology.
       \item Rescaling in apparent horizon radius also alters the Cai--Kim temperature. 
       \item The fractal density parameter of the fluid is modified by a prefactor which depends on the fractal dimension, Hubble parameter, and the grid size.
\end{itemize}

Our encountered equations show that dark energy is required to explain the Universe's present stage and associated observations. It is noteworthy that when $D=3$, the standard Friedmann equations are restored. Our model offers an alternate interpretation, speculating that the measurable parameters could have been affected by fractal properties. In particular, changing the fractal dimension may have a major impact on the age of the Universe as well as the deceleration and transition redshift at the current epoch.

In conclusion, we have only just begun to explore and refine our fractal model, which is a modified cosmology. We contribute to the advancement of fractional models and may find exciting new findings at both quantum and classical levels by examining a number of modern cosmological issues, including the Hubble tension, gravitational wave propagation,  gravitational lensing, and density perturbation generation. However, it is prudent to defer the above evaluations to later undertakings.

\section*{Acknowledgements}
S.J. acknowledges financial support from the National Council for Scientific and Technological Development--CNPq, Grant no. 308131/2022-3. 
\bibliographystyle{elsarticle-num}

\bibliography{Apparent}

\end{document}